\documentclass[12pt]{article}

\usepackage{graphicx}
\usepackage{amssymb}
\usepackage{psfrag}
\usepackage{color}

\renewcommand{\=}{~=~}

\begin{document}

\begin{center}
   \textbf{\Large HOLLYWOOD BLOCKBUSTERS\\[1mm] Unlimited Fun but Limited Science Literacy}\\[2mm]
  C.J. EFTHIMIOU\footnote{costas@physics.ucf.edu} and
  R.A. LLEWELLYN\footnote{ral@physics.ucf.edu}\\
  DEPARTMENT OF PHYSICS\\
  UNIVERSITY OF CENTRAL FLORIDA
\end{center}

\section{Introduction}

There is no doubt that Hollywood has become an established  major
source for entertainment in the lives of the citizens of the
modern society. In the products of Hollywood (big screen movies,
TV mini series, TV series, sitcoms, etc) amazing feats are
presented by people supposedly the best in their fields. Great
scientists find solutions to major scientific challenges, the best
NASA employees save the Earth from the ultimate heavenly threats,
the best soldiers defeat armies on their own, the best psychics
solve criminal cases, the best parapsychologists manage to
successfully investigate supernatural phenomena and so on. And of
course we should not forget the laypersons who often save the day
by finding solutions that scientists could not think of.
Unfortunately all this is only great entertainment. When
logic and science are used to decide if certain scenarios are
consistent and plausible, usually the results are disappointing.
The inconsistencies of the Hollywood products with science may come
as a surprise to many people who simply accept what they see as
realistic or, at worst, slightly modified from reality.

In this article, we will examine specific scenes from popular
action and sci-fi movies and show how blatantly they break the
laws of physics, all in the name of entertainment, but
coincidentally contributing to science illiteracy. Towards this
goal, we assume that our reader has an understanding of
algebra-based general physics.


\section{Cinema Fermi Problems}

Fermi problems (also known as back-of-the-envelope problems)
\cite{Baeyer} have been very popular among physicists
\cite{Swartz} since Fermi used them to illustrate his dramatic and
extraordinary ability to give approximate answers to the most
esoteric and puzzling questions. In a simple adaptation of the
idea, we have applied it to plots and particular events appearing
in Hollywood movies \cite{EL,EL3} to help us decide the
plausibility of the plot or the event. Often such an analysis is
not necessary because the impossibility of the action can be
explained qualitatively. Such scenes are those presented in
sections \ref{sec:Speed}, \ref{sec:Spiderman}, \ref{sec:Aeonflux},
\ref{sec:Core}, \ref{sec:Superman}. However, some simple
calculations reveal additional absurdities.

\subsection{Ignorance of Projectile Motion}
 \label{sec:Speed}

    In the movie \emph{Speed} \cite{Speed} a bus that has been booby-trapped should not
    drop its speed below 50 mph,
    otherwise a bomb will explode killing everyone on board. As the bus is moving on a highway,
    the people on the bus are informed that, due to road construction, a bridge in the highway
    is missing
    its center segment. Unable to stop the bus, the decision is made to jump over the gap.
    The bus then accelerates to almost 70 mph and, of course, successfully makes the
    jump\footnote{Time: 1:05:03--1:06:41}.

    \begin{figure}[h!]
    \begin{center}
    \includegraphics[width=7cm]{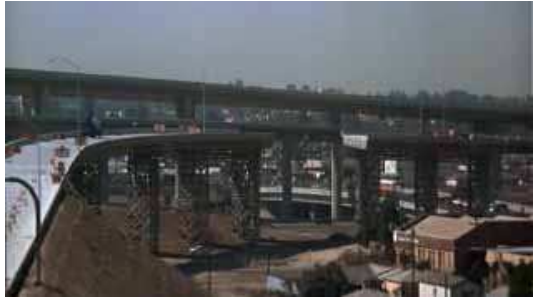}
    \end{center}
    \caption{The gap in the highway in the movie \emph{Speed}. Notice that the bridge
             is perfectly horizontal.}
    \end{figure}

    The movie gives us several shots of the gap in the highway.
    The viewer can clearly see that the highway is level at the
    bridge. Unfortunately, this predetermines the destiny of the
    bus: there is no way that it will jump over the gap. As soon
    as it encounters the gap, the bus will dive nose down to hit
    the ground below the bridge.

    \begin{figure}[h!]
    \begin{center}
    \includegraphics[width=7cm]{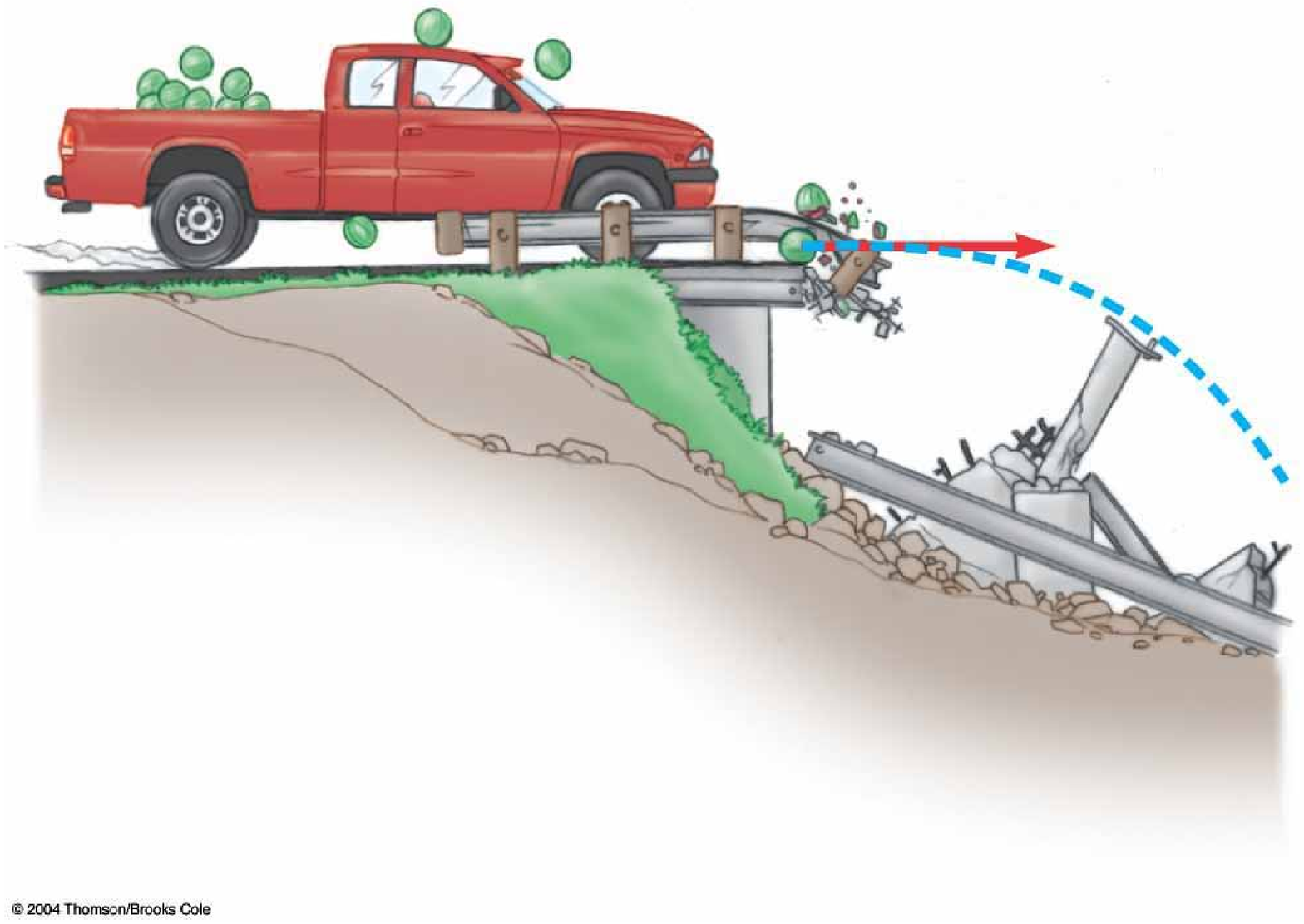}
    \vspace{5mm}
    \includegraphics[width=7cm]{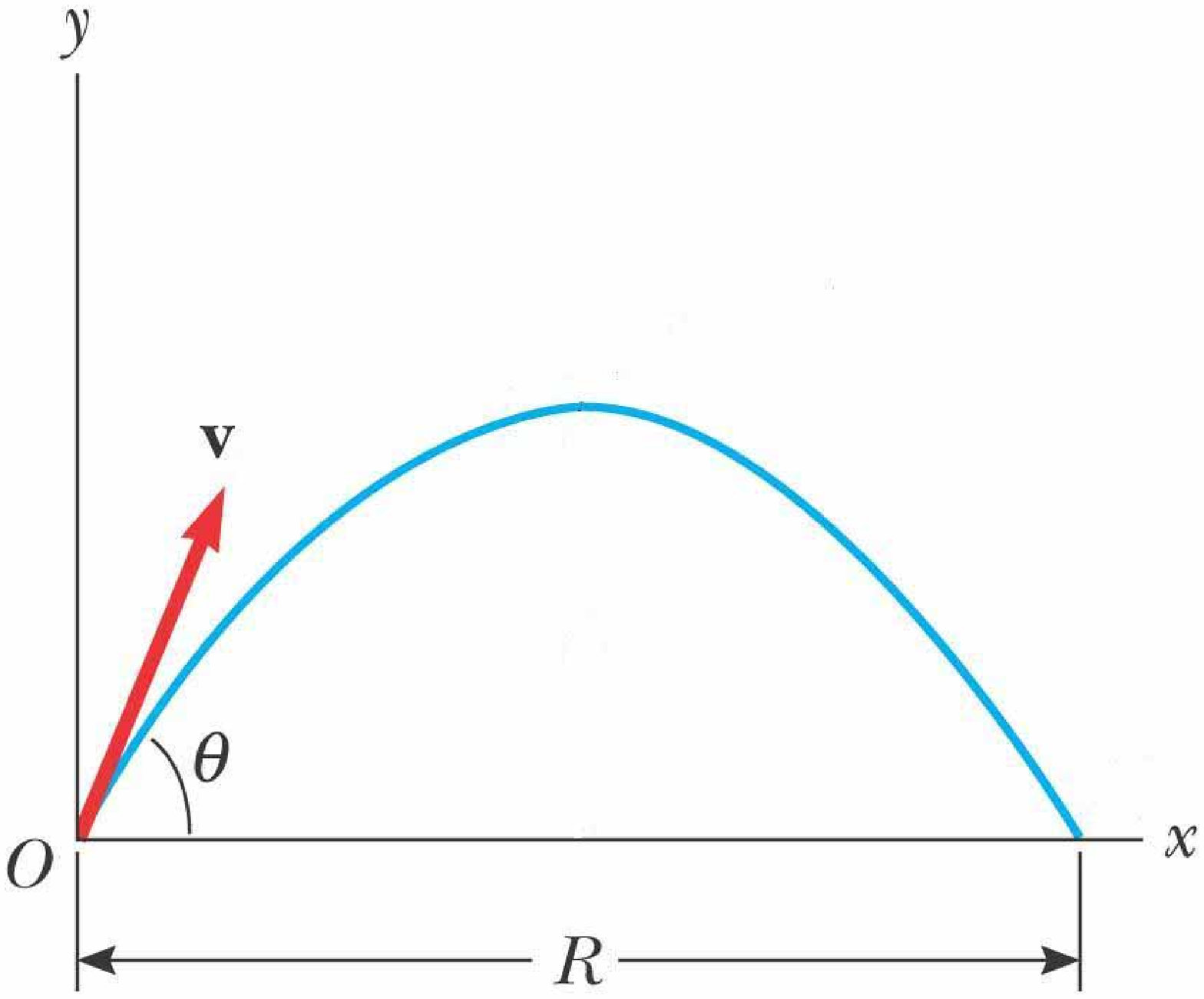}
    \end{center}
    \caption{Left:
             A car (or bus) going over a bridge gap with horizontal initial speed will dive
             nose down the bridge as soon as it is over the gap.
             (Picture from \cite{Serway}).
             Right:
             If the initial velocity of the car (not the car!) has a tilt $\theta$,
             the car will follow
             a parabolic path that, depending on the magnitude of the velocity and the tilt, may
             be long enough to allow the car reach the other side of the bridge.}
    \end{figure}

    At least, the director and the special effects team seem aware of the above
    fact. So, upon looking carefully at the scene, we see the bus
    depart from the highway at an angle of about $30^\circ$ relative to the horizontal.
    Of course, this is evidence of a miracle as it would happen
    only if a ramp had been placed exactly before the gap. In the
    movie, as the protagonists talk to each other, a laughable explanation is given:
    `the road leading to the bridge is uphill'.

    \begin{figure}[h!]
    \begin{center}
    \includegraphics[width=4cm]{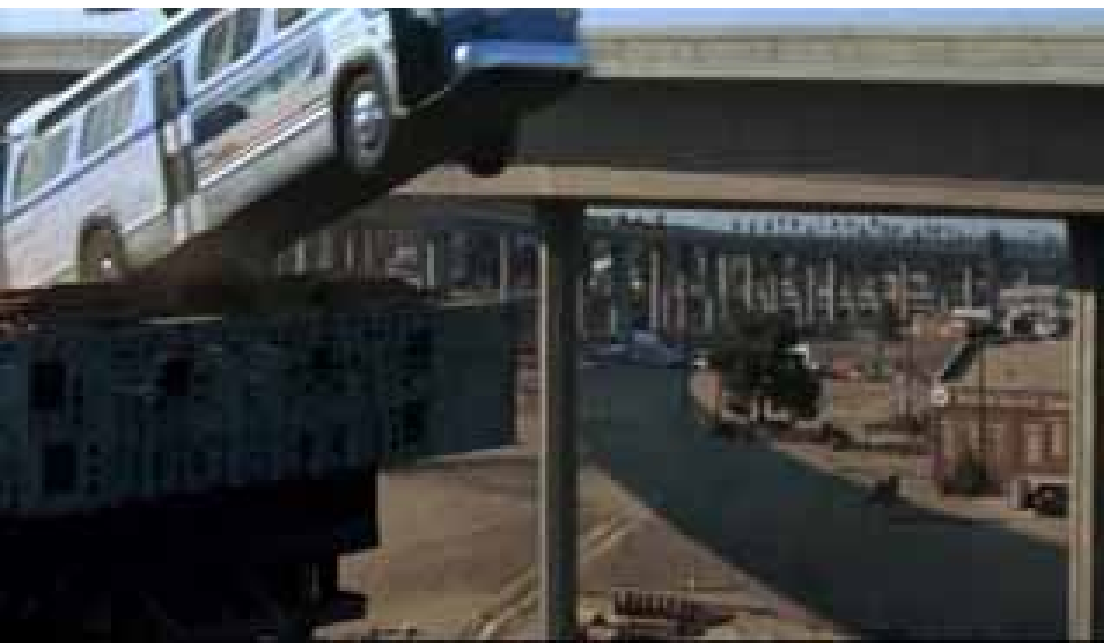}
    \includegraphics[width=4cm]{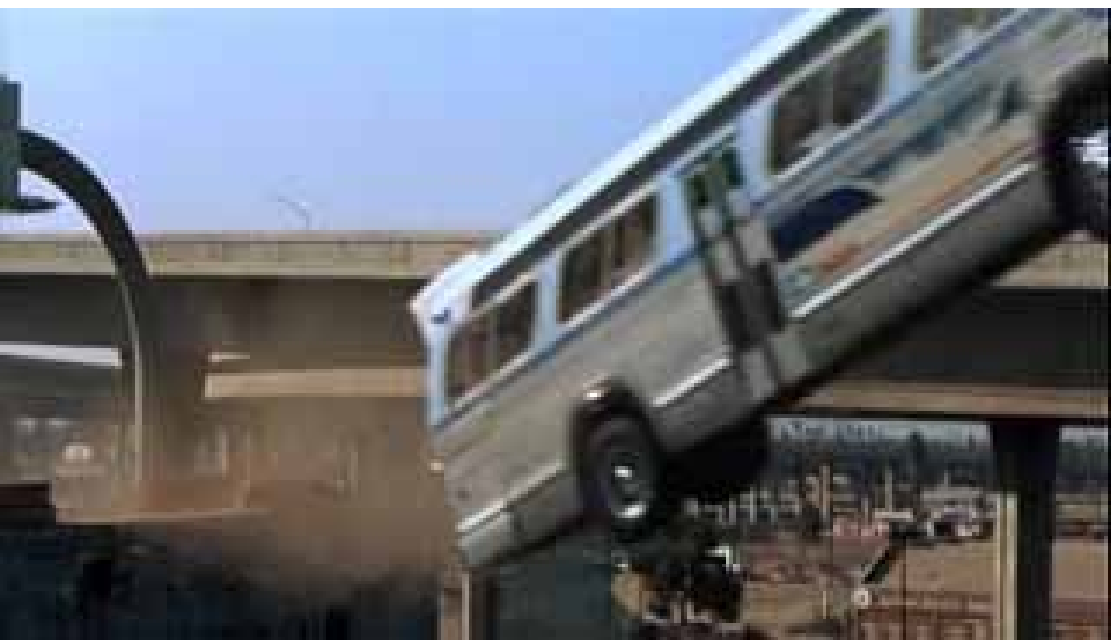}
    \includegraphics[width=4cm]{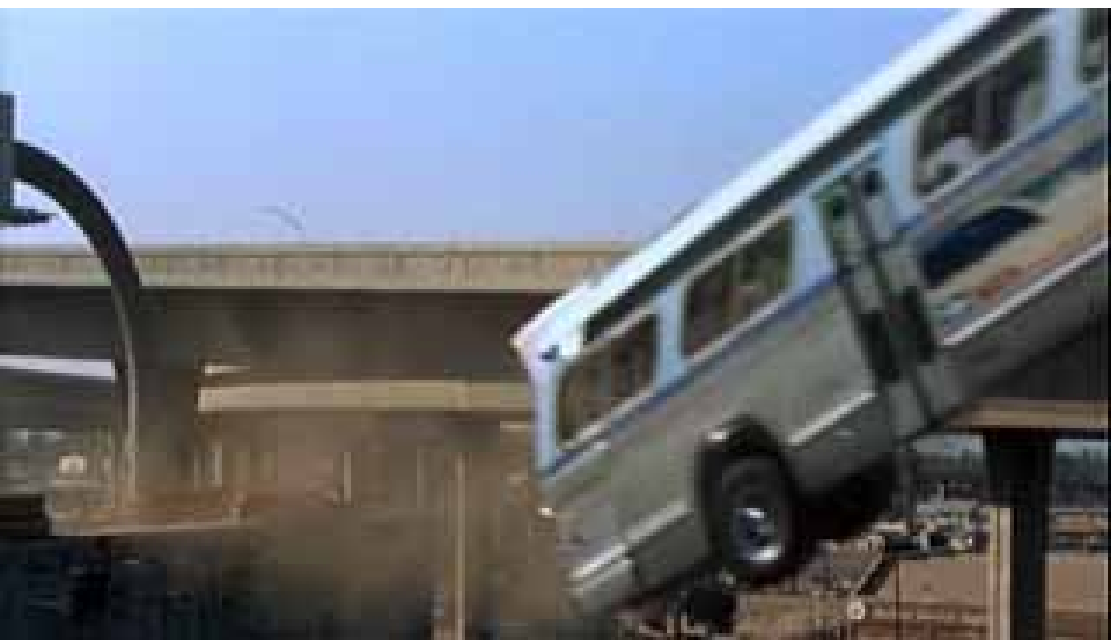}
    \includegraphics[width=4cm]{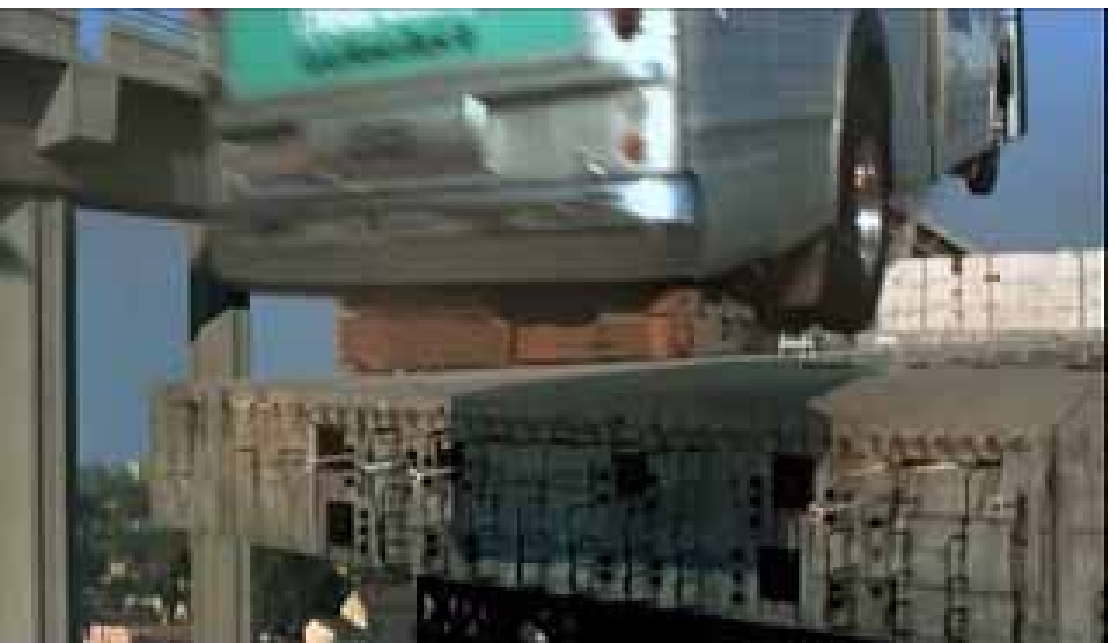}
    \end{center}
    \caption{A sequence of stills as the bus in \emph{Speed} jumps over the bridge gap.}
    \end{figure}

    In any case, given the miracle, the scene is still
    problematic. Paying attention to the details of the scene, it
    looks as if the back end of the bus drops a little after it
    is over the gap. Probably this is not something the
    director wanted to show; it may be a remaining flaw from the
    special effect used to create the scene. However, there seems to
    be another serious problem: the director shows that, although the
    bus has tilted upward at an angle, it then flies over the gap in a
    straight horizontal line! Unfortunately, it is not very easy to
    verify the trajectory of the bus as
    the director does not show the whole jump; there are hints in
    the scene pointing to either interpretation\footnote{Just watch the clip in slow motion
    carefully to reach your own conclusion.}: an incorrect horizontal trajectory and a curved
    trajectory.

    Of course, the jump over the bridge is an example of  projectile motion
    with initial speed $v$ and initial
    angle $\theta$. The bus's path, like any projectile, must be a parabolic one with
    its peak at the middle of the gap if the speed and angle are such that the bus
    will just make it over the gap. If the the initial speed and
    angle are more than enough, then the peak of the path may be
    shifted towards the the right.

    Ignoring frictional and drag forces, for a projectile motion
    the range would be
    $$
       R \= {v^2 \, \sin(2\theta)\over g}~.
    $$
    Given the movie data (angle $\theta=30^\circ$, speed $v=70 mph=31 m/s$) and
    that $g=9.8m/s^2$, this formula implies a
    range of 85.5 meters. Since the situation seen in the movie
    must include frictional and drag forces, we may
    approximate roughly the range of the bus at full speed
    at  40 meters or about 131 feet. This is less than half the
    ideal range; usually the range will not be reduced so
    drastically.
    So, given the miracle that not only the bus will tilt but,
    also the velocity vector will tilt at an initial angle of 30 degrees,
    the bus can jump more than 130 feet.
    However, the gap is only 50 feet as we are told in the movie. So, the bus
    should have landed much further on the other side, at least the length of the bus beyond
    the edge, and not close to the edge of the gap as shown.

\subsection{Ignorance of Newton's Laws}
 \label{sec:Spiderman}

    In Spiderman \cite{Spiderman} the villain Green Goblin
    kidnaps Spiderman's girl friend Mary Jane (M.J.) and takes her  on
    the tower of the Queensboro bridge. There, while waiting for
    Spiderman, he cuts loose the cable that supports the tramway
    cabins, which commute between Manhattan and Roosevelt island, and takes hostage
    a tramway cabin  that is full of children. When Spiderman shows up,
    the Green Goblin  is holding in one hand the cable that supports the cabin with the
    children and M.J. in the other hand\footnote{Time: 1:39:34--1:42:10}.

    \begin{figure}[h!]
    \begin{center}
    \includegraphics[width=7cm]{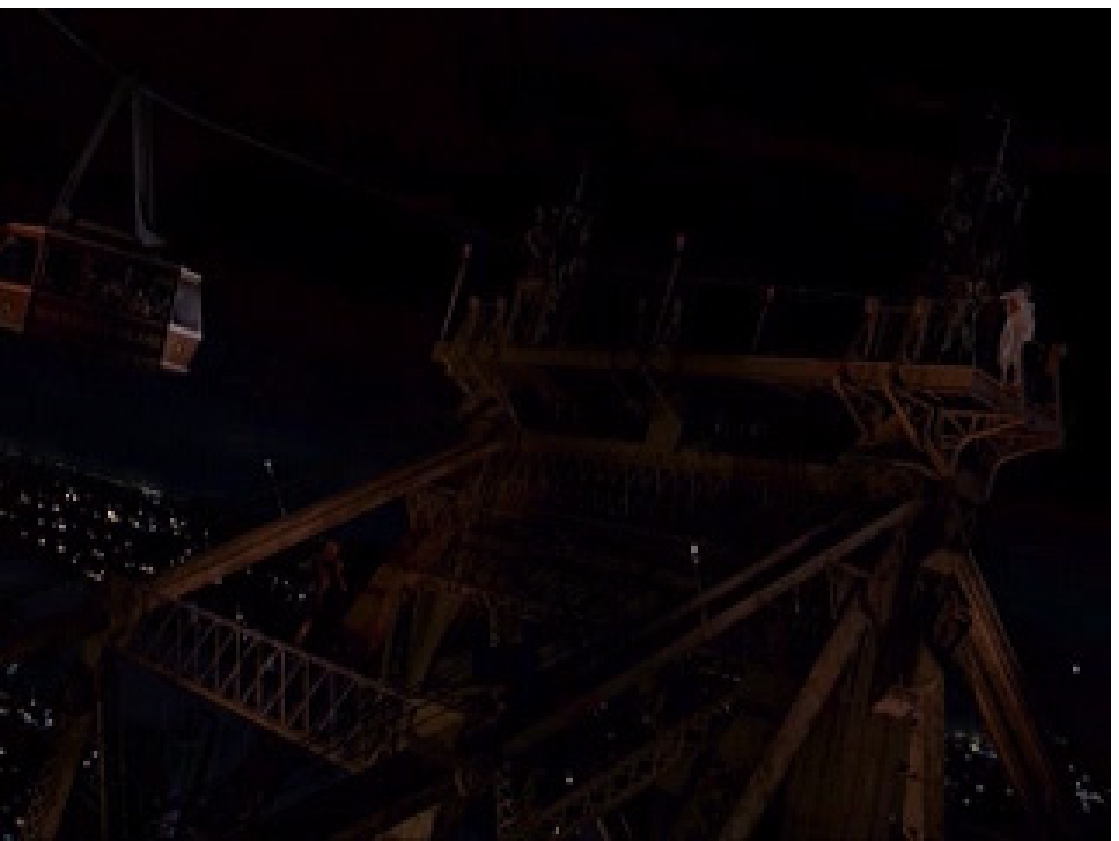}
    \includegraphics[width=7cm]{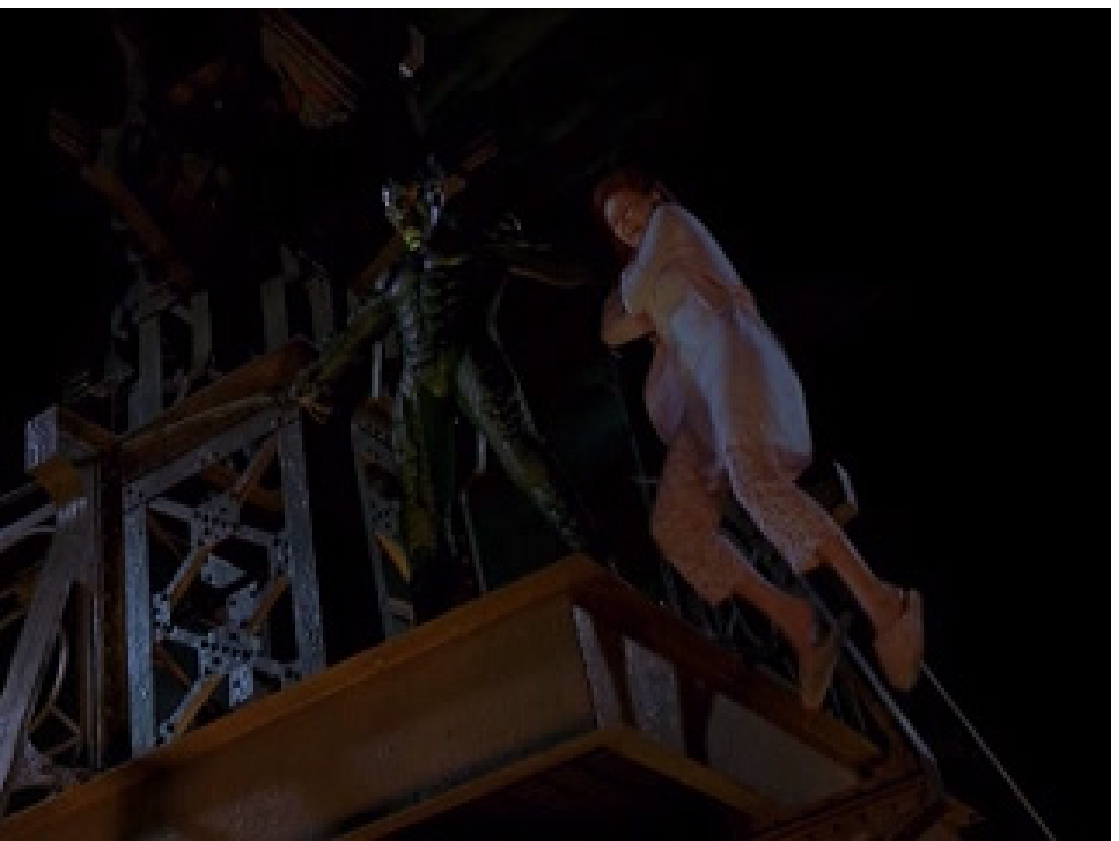}
    \end{center}
    \caption{Left: Green Goblin in static equilibrium while he holds M.J. and the cabin.
             Right: A close-up of Green Goblin's static equilibrium position.}
    \label{fig:Spiderman}
    \end{figure}

\begin{figure}[h!]
    \begin{center}
    \includegraphics[width=7cm]{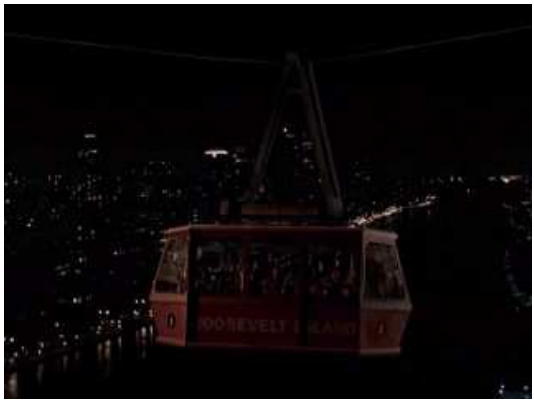}
    \end{center}
    \caption{Close-up of the trapped cabin with the children.}
    \label{fig:Spiderman2}
    \end{figure}

    There are some problems with this scene (and its continuation
    as shown in the movie). Notice in the left still of figure
    \ref{fig:Spiderman} that the cable has the shape of a nice
    smooth curve even at the point where the car is located. If a
    heavy object is hung from a flexible rope, then at the
    point of the rope where the object is attached we should see a
    `kink'---that is, a sharp point where the curve is not smooth
    anymore. However, at another close-up view, the director does
    show the kink. (Figure \ref{fig:Spiderman2}.) It is possible
    that    the kink in the still of figure \ref{fig:Spiderman} is hidden
    due to the angle the still is taken (so we will not hold this against the director).

    In the left still of figure \ref{fig:Spiderman}, it appears  that the
    left end of the cable is anchored at a higher location
    relative to the position of the Green Goblin (who is standing on the top
    of the bridge tower). This would imply that the cabin should slide down the cable
    towards the Green Goblin. However, in the still of figure
    \ref{fig:Spiderman2}, it appears that the two ends of the cable are at the same height.
    Again, we may assume that the illusion in the first still might be due to the angle the
    still has been taken. On the other hand, if we look at the
    construction data for the bridge \cite{Queensboro} and the Roosevelt Island
    tramway \cite{RIOC}, we discover that a stretched cable between the top of the bridge tower
    and the tramway towers cannot be horizontal. In fact, Green Goblin is located at a much higher point.
    The height of the bridge tower above water is 350 feet while  the
    tramway, at its highest point, is 250 feet above the water.

     \begin{figure}[h!]
     \color{red}
    \begin{center}
    \psfrag{a}{$F$}
    \psfrag{b}{$F$}
    \psfrag{c}{$W$}
    \psfrag{t}{$\theta$}
    \includegraphics[width=7.2cm]{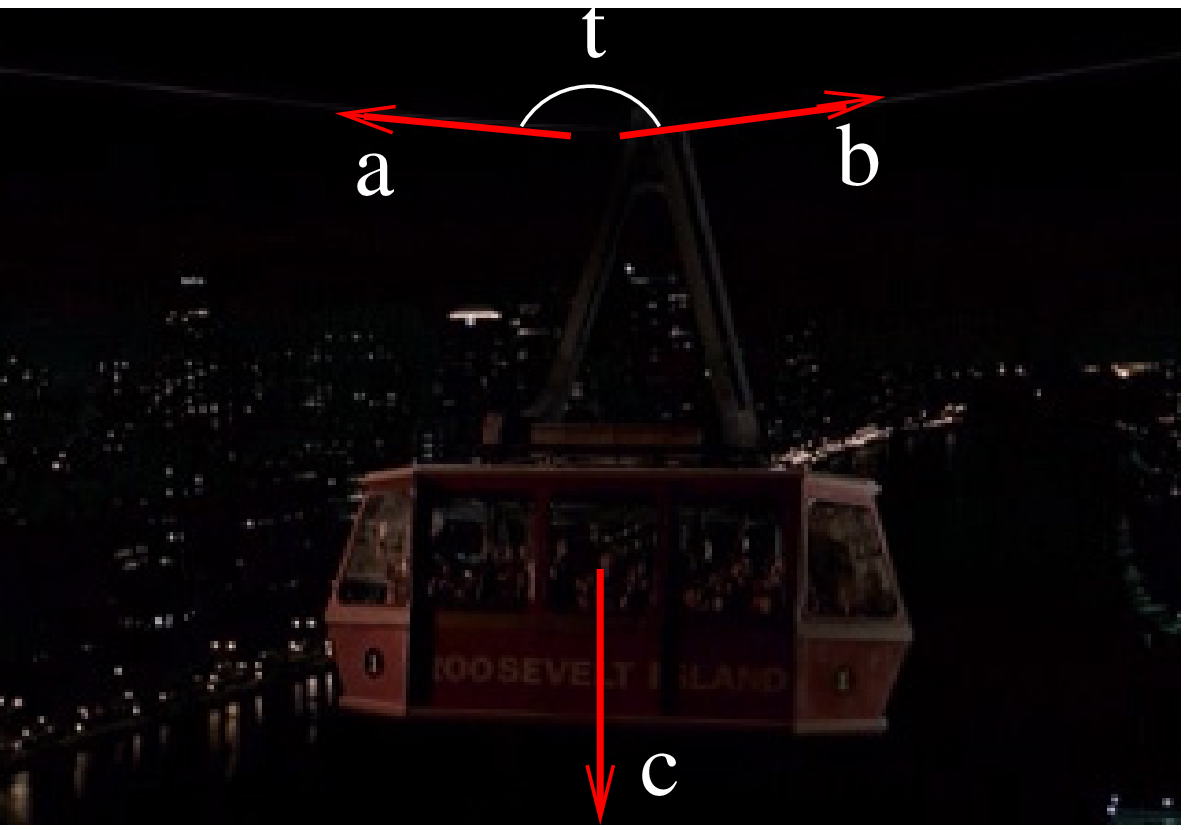}
    \psfrag{a}{$F$}
    \psfrag{b}{\hspace{-2mm}$N_1$}
    \psfrag{c}{\hspace{-1mm}$Mg$}
    \psfrag{d}{\hspace{-2mm}\raisebox{3mm}{$N_2$}}
    \psfrag{e}{$mg$}
    \includegraphics[width=7cm]{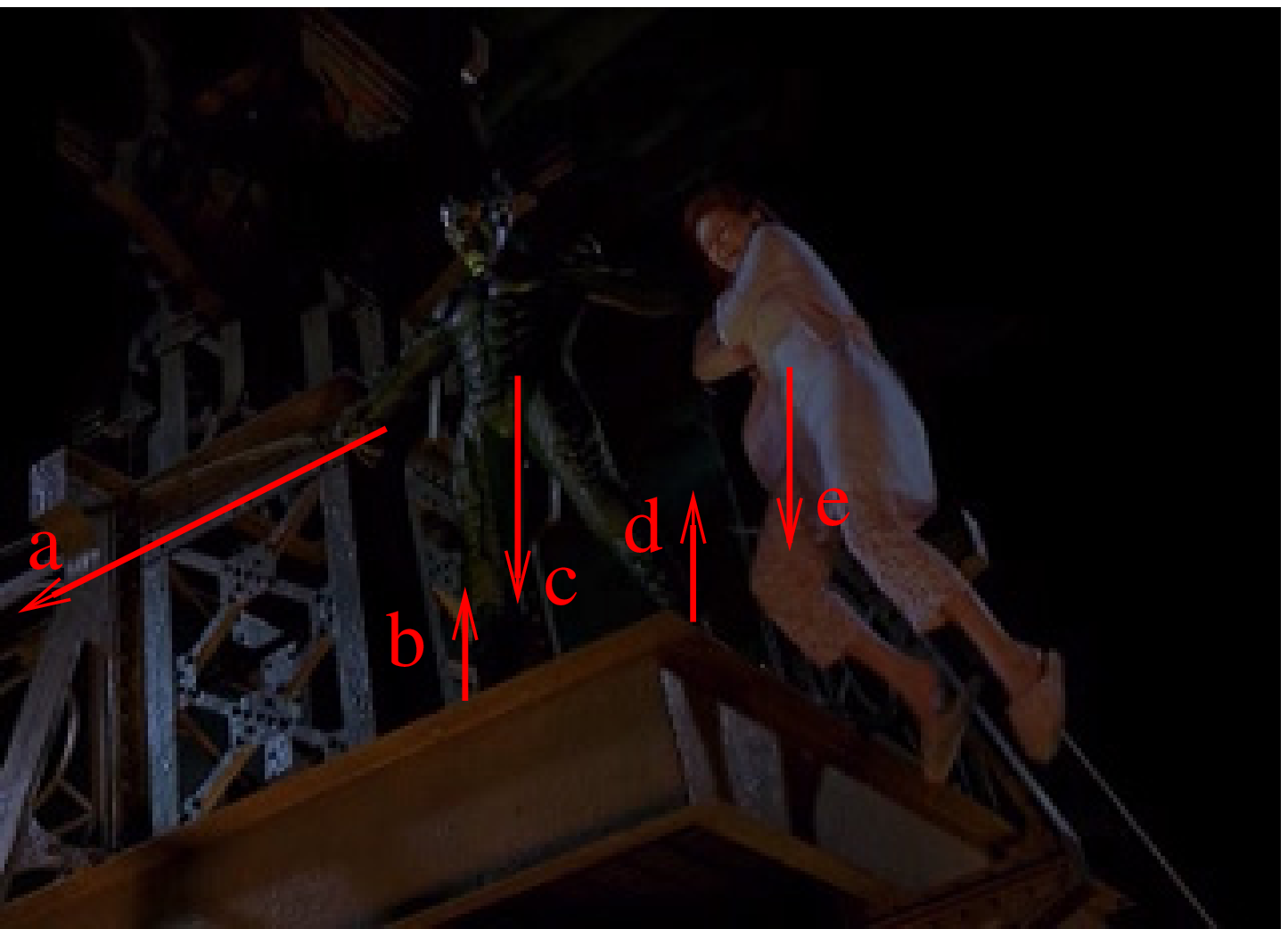}
    \end{center}
    \color{black}
    \caption{The relevant forces for the scene. Left: The free body diagram for the cabin.
             Right: The free body diagram for the Green Goblin. The sum on the two normal
             forces $N_1$ and $N_2$ at the feet of the Green Goblin is equal to the total
             normal force $N$ as used in the text.}
    \label{fig:Spiderman3}
    \end{figure}

    For the present purposes, we will ignore these technicalities and assume that the
    two ends of the cable are indeed at the same height.
    Furthermore, to simplify the math, we shall assume (although not an essential
    assumption in the calculation) that the
    cabin has been trapped at the midpoint of the cable. The
    latter assumption implies that the two forces $\vec F_1$ and $\vec F_2$
    from the cable on the cabin
    (see figure \ref{fig:Spiderman2}) are equal in magnitude---say $F$.
    From  figure \ref{fig:Spiderman2}, we see that the angle the cable makes with the
    horizontal is  $\theta=7^\circ$. Then
    $$
         2F\,\sin\theta \= W~,
    $$
    where $\vec W$ is the total weight of the cabin.

    The following forces are acting on the Green Goblin:
     (a) a force $\vec F_3$ from the cable. Since the cable is in
        equilibrium, the tension along its length is equal to $F$
        and thus this is the magnitude of $\vec F_3$;
     (b) a downward force $m\vec g$ equal to M.J.'s weight that is acting on his left
        hand;
     (c) his weight $M\vec g$;
     (d) the normal force $\vec N$ from the tower pointing upwards.
    In the vertical direction, the forces $+mg$, $+Mg$, $-N$, and the
    component $+F\sin\theta$ of $\vec F_3$ cancel out:
    $$
        mg + Mg + F\sin\theta- N \= 0~.
    $$
    However, there is no force to cancel the horizontal component
    $F\cos\theta$
    of $\vec F_3$. This implies that the Green Goblin---no matter
    how
    strong he is---cannot stay in static equilibrium.

    One can try to save the situation by claiming that a
    static frictional force $\vec f_s$ must also be in operation. Indeed, in
    this case the forces seem to cancel in the horizontal direction
    too:
    $$
             F\cos\theta - f_s \= 0~.
    $$
    However, Green Goblin still cannot stay in static
    equilibrium. Given that the maximum value of static friction
    is $\mu_s N$, cancellation of the forces requires that
    $$
       \mu_s ~\ge~ {W\over 2\tan\theta\,(W/2+Mg+mg)}~.
    $$
    The cabin's weight is much bigger than the combined weight of
    M.J. and the Green Goblin. So, $W/2+Mg+mg\simeq W/2$ and
    therefore
    $$
       \mu_s ~\ge~ {\cot\theta\over2} ~\simeq~ 4~.
    $$
    Coefficients of friction are usually below 1. In some
    exceptional cases they can be higher than 1 but a coefficient
    of 4 is extremely high and probably attainable only if the
    materials in contact have adhesive properties.
    Besides the fact that it is not easy to obtain the high value of friction necessary,
    even if we did have it, the two forces $\vec F_3$ and $\vec f_s$ would act at different
    locations and
     would create
    considerable torque that could not be matched by the opposing
    torque created by $m\vec g$.

\subsection{Ignorance of Impulse}
\label{sec:Aeonflux}

\begin{figure}[h!]
    \begin{center}
    \includegraphics[width=7cm]{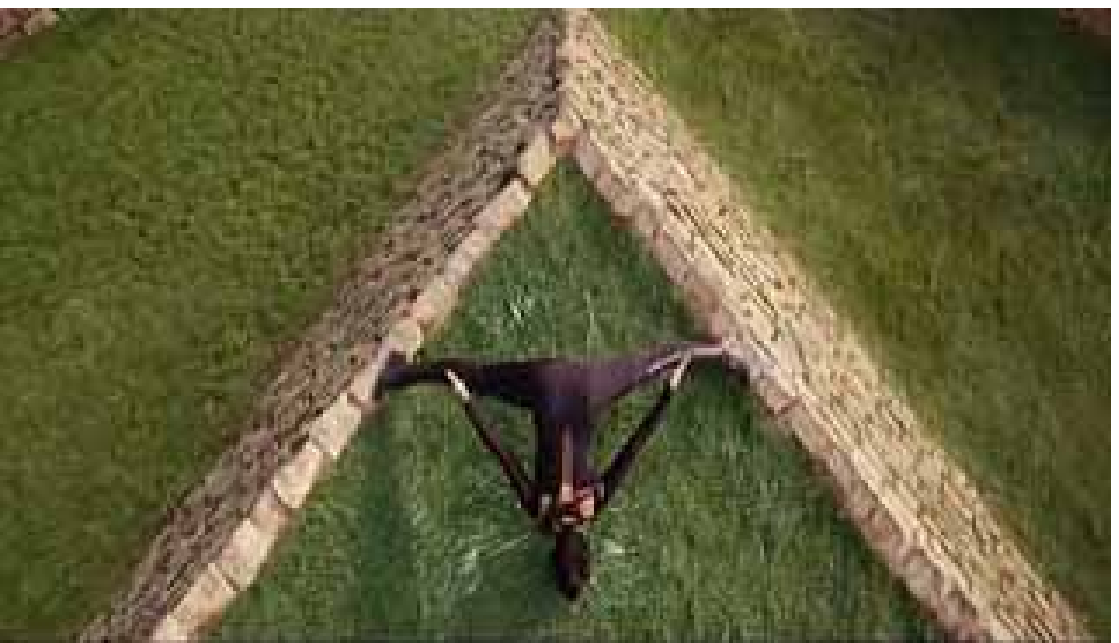}
    \includegraphics[width=7cm]{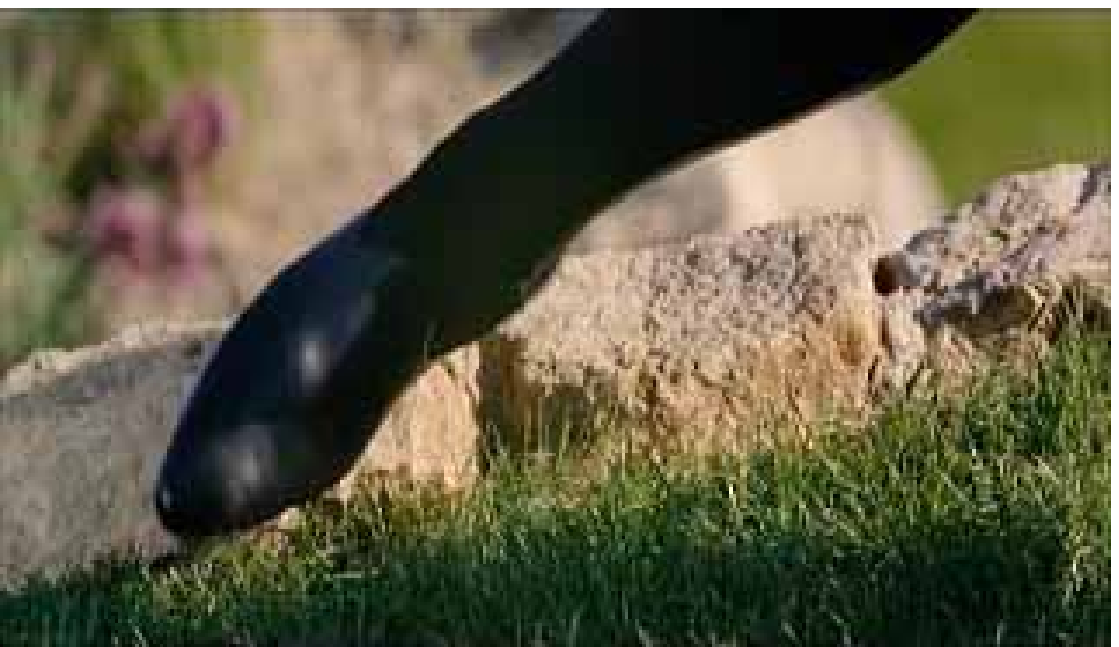}
    \end{center}
    \caption{Left: Aeon Flux's life saver landing position. Right:
             Aeon Flux's feet at the previous position.}
    \label{fig:Aeonflux}
    \end{figure}

    Aeon Flux is a rebel assassin with superhero capabilities.
    She is working for the Monicans, a group of rebels trying to overthrow the government.
    She is  sent on a mission to kill the Chairman, the head of the government.
    Assisted by Sithandra, Aeon is trying to reach the government's building that is surrounded by
    a booby trapped field. In an effort to defeat the defensive system that monitors the field,
    Aeon uses her gymnastic abilities. She displays a series of cartwheels and summersaults.
    In one of such display, as she lands she notices that sharp blades are coming out of the
    grass. To compensate, as soon as she lands her feet on the stone boarder of the
    grass, she stops her forward movement by arranging her body in
    the position shown in the left picture of figure \ref{fig:Aeonflux}. Although her body comes
    close to the blades, she never touches them, thus escaping a fatal encounter\footnote{Time:
    0:17:45--0:18:08}.

     All this might  excite the audience of the movie and, and especially   the sympathizers
     of the heroine, but the scene is ridiculous.  As Aeon Flux lands,
     she has forward (and downward) momentum. To change it, she
     needs to be acted upon by an external impulse. The ground
     certainly can stop her downward motion. However, there is
     nothing that can stop her forward momentum---well, except the
     frictional force at the feet of Aeon. If such frictional
     forces exist, they create torque that reinforces the torque
     created by Aeon's weight forcing her to rotate and drop on
     the sharp blades...

     Even if Aeon could magically stop her forward motion and
     place herself in the position shown in the left still of figure
     \ref{fig:Aeonflux}, this position is not an equilibrium position at
     all. As long as her center of mass is not above the line defined
     by her two feet, the torque created will force her to drop
     on the grass. Her center of mass is somewhere in her waist;
     it is evident from the figure that it is certainly not above
     the line joining her feet.

     In an effort to exaggerate the abilities of Aeon, the
     director makes things worse for himself. He shows to us a close-up view of the way
     Aeon is standing on the stones. The stones defining the border
     of the grass are cut with a slope---each stone looks similar to an inclined plane. At the
     same time the stones have been placed to make a V-shaped
     border. This configuration makes obtaining equilibrium a really
     difficult task.

    \begin{figure}[h!]
    \begin{center}
    \includegraphics[width=7cm]{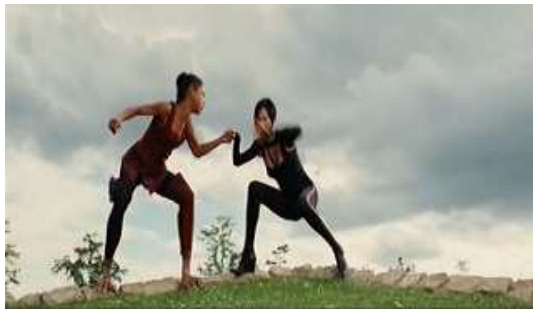}
    \end{center}
    \caption{Sithandra helps Aeon restore her balance.
                   Look at the stones at the lower right side to see clearly the way they are
                   cut.}
    \label{fig:Aeonflux2}
    \end{figure}

     In order to be fair, we should make the following comment.
     Internal forces, although they always add to zero, can create
     a net torque. So, in principle, Aeon's muscles can create the
     necessary torque to stop her fall to the grass. Since she is
     a superhero, we can imagine that she has this ability\footnote{Of course, if this is the
     case, it is not easy to explain why she needs Sithandra's help to rotate her body back to the
     vertical position as her muscles can push her back. One must assume that
     the position she has acquired uses the maximum torque her muscles can
     generate. But then, as she lands she cannot stop her fall since she has initial angular momentum
     due to her motion and thus her muscle needs to exceed the amount of torque they apply
     on equilibrium... Well, as our students say, stop thinking...}.
     It is thus not the way she stands
     still, but the way she flexes her muscles that keeps her in equilibrium.

\subsection{Ignorance of Buoyancy}
\label{sec:Core}

Hollywood has produced many silly movies whose plot does not make
any sense. But only few of them are as awful as \emph{The Core}
\cite{Core}. Due to military experiments, the outer core of Earth
has stopped rotating thus leading to a dramatic drop in Earth's
magnetic field that protects Earth from harmful radiation. A team
of gifted scientists and pilots is assembled to drive an
innovative subterranean vehicle to the core so that, with the help
of a nuclear bomb, they may restore its rotation. As the team
descends, at a depth of 700 miles, the vehicle crashes in an
underground cave. After repairing the damage, one member of the
team is killed by flying debris and his dead body drops in lava
where it sinks in few seconds\footnote{Time: 1:09:58--1:10:45}.

    \begin{figure}[h!]
    \begin{center}
    \includegraphics[width=4cm]{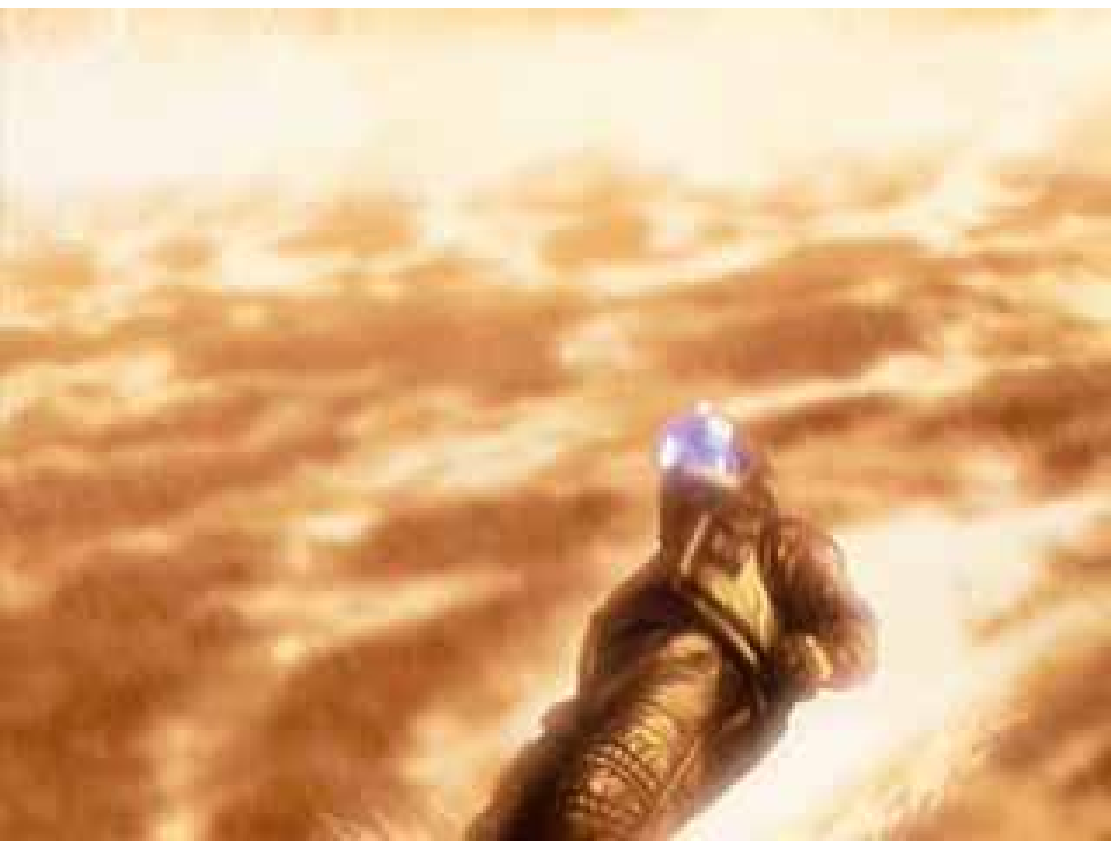}
    \includegraphics[width=4cm]{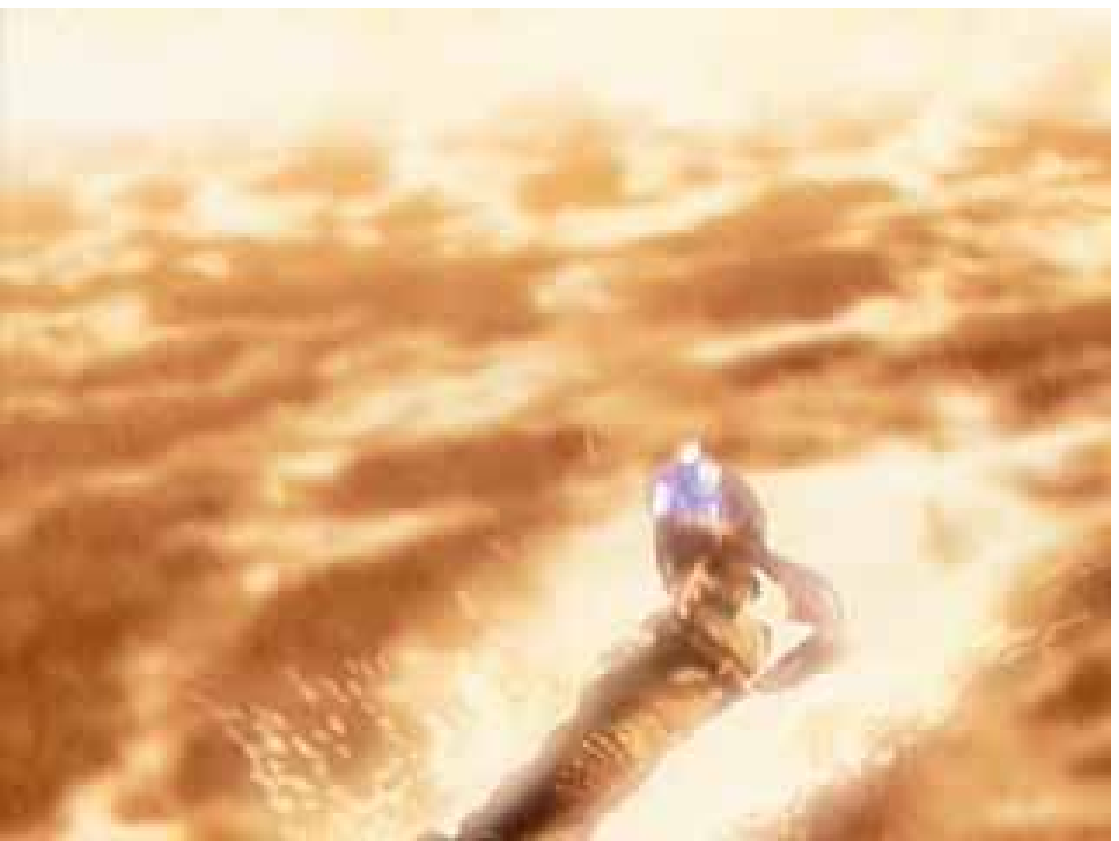}
    \includegraphics[width=4cm]{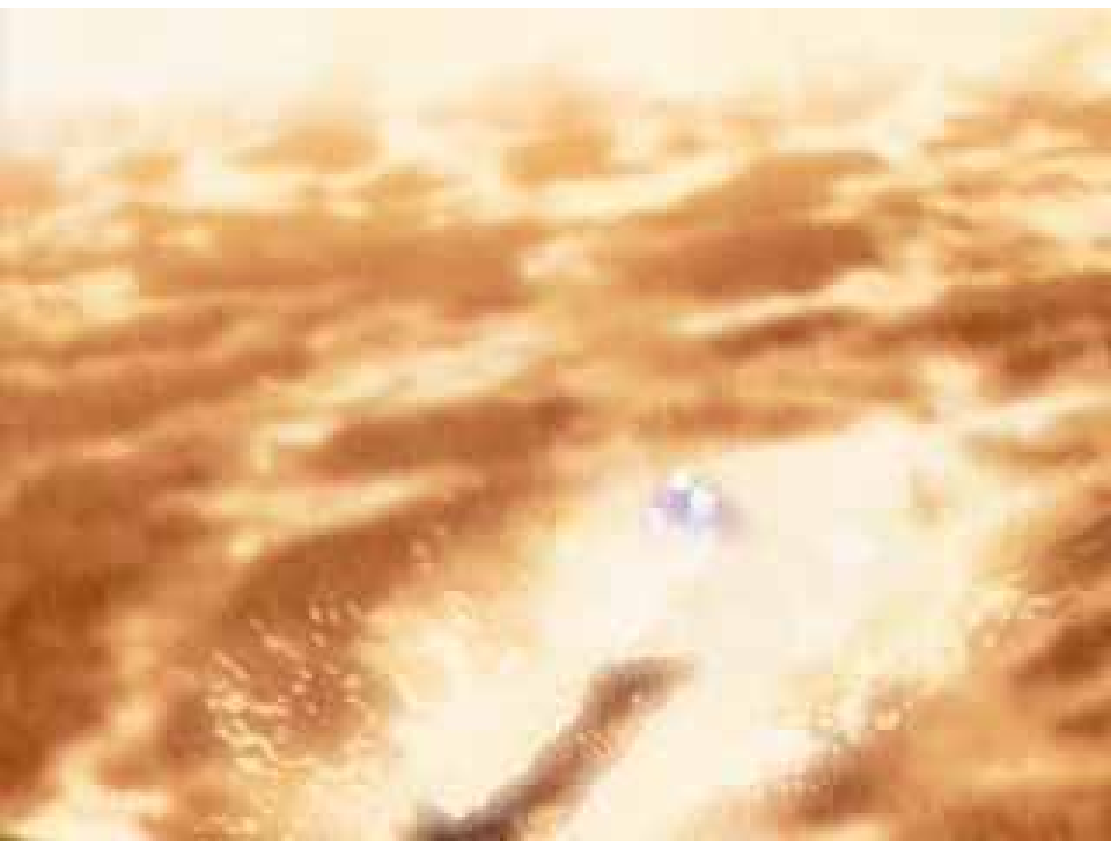}
    \end{center}
    \caption{A sequence of stills as a body sinks in lava.}
    \label{fig:lava}
    \end{figure}

We could discuss many questionable issues with the scene: (a)
Could a cave  exist in such depths? (b) Could the crew  afford to
open and close the door of the vehicle in such a depth? This would
mean loss of breathable air from the vehicle and changes in the
air pressure and temperature of the vehicle. (c) Could the
flexible suits that the crew is wearing really protect them at
that depth?  Many more questions could be added in this list. The
reader can reflect on these issues on his own. We will only
discuss the sinking of the human body in the lava\footnote{For the
interested reader we point out that an additional cinema Fermi
problem based on the plot of \emph{The Core} can be found in
\cite{EL3}.}.

Let's study what happens to an object that is thrown to the pond
of lava. Two forces are acting on the object. First, there will be
the force of gravity acting on the object:
$$
   F_{gravity} \= g_{below}\, M_{object}
$$
where $M_{object}$ is the mass of the object and $g_{below}$ is
the acceleration of gravity at the location of interest. The
latter is actually weaker than that on the surface of Earth
since---according to a well known theorem of Newtonian
gravity---only the material of the Earth located in the sphere
below the current depth will contribute to the gravitational
attraction:
$$
    g_{below} \= G \, {M_{below}\over R_{below}^2}~,
$$
where $M_{below}$ is the mass of the Earth contained in the
corresponding sphere. Assuming that the Earth has uniform density,
$M_{below}={4\over3}\pi R^3_{below} \, \rho_{Earth}$. Therefore
$$
   g_{below} \= {4\pi G \over 3} \, \rho_{Earth} \, R_{below}~.
$$
In the same way, the acceleration of gravity $g=9.8m/s^2$ at the
surface of the Earth is given by
$$
  g \= {4\pi G \over 3} \, \rho_{Earth} \, R_{Earth}~.
$$
Dividing the last two equations, we find that
$$
  g_{below} \= g \, {R_{below}\over R_{Earth}}~.
$$
All this implies that as the vehicle moves closer to the center of
the Earth, the force of gravity on it weakens. At the crash site
it is already 20\% less. This conclusion has some serious
implications;  the people could not walk and move normally
(contrary to what we see in the movie). In any case, ignoring such
implications, we will only look at the sinking of the body.

Using the volume $V$ and density of the object $\rho_{object}$, we
can now write
$$
  F_{gravity} \=  {R_{below}\over R_{Earth}}\, g\, \rho_{object} \,
  V~.
$$
Besides gravity, once an object is inside the lava, there is also
the force of buoyancy from the lava.
 This is equal to the gravitational force felt by the displaced
 lava:
$$
  F_b \= g_{below} M_{displaced}
      \= {R_{below}\over R_{Earth}}\, g\, \rho_{lava} \, V_b~,
$$
where $V_b$ is the volume of the object that is submerged in the
lava. The net force on the body is thus
$$
  F_{net} \= {R_{below}\over R_{Earth}}\, g\, (\rho_{lava} \, V_b-\rho_{object} \,
  V)~.
$$
Notice that, depending on the sign of the quantity
$$
  \rho_{lava} \, V_b-\rho_{object} \, V~,
$$
an object can float or sink.

The human body is made mainly of water, thus its density will be
almost equal to that of water, $\rho_{water}=1000kg/m^3$. The lava
is mostly molten rock; surface rocks have an approximate density
of $3300kg/m^3$. So $\rho_{lava}=3300kg/m^3$. Therefore, for the
human body,  once a third of it submerges in lava, the two forces
become equal and the body stops sinking. Even more, sinking (in
lava) will happen at a slower rate compared to the rate on the
surface of the Earth since gravity is weaker at that depth.

\subsection{Ignorance of Angular Momentum and More}
\label{sec:Superman}

In the movie \emph{Superman} \cite{Superman}, Superman, being
unable to stand the loss of his great love Lois Lane,  decides to
reverse the rotation of the Earth so he can reverse time. He thus
flies very high---outside the Earth's atmosphere---and starts
revolving around Earth at a great speed. After doing so for some
time, the Earth finally slows down and then starts rotating in the
opposite direction. This forces time to run backwards bringing the
clock before the death of Lois. Once he succeeds in `resurrecting'
Lois, he changes his direction of revolution  eventually forcing
Earth to return to its original direction and rate of
rotation\footnote{Time: 2:18:34--2:20:13}.

    \begin{figure}[h!]
    \begin{center}
    \includegraphics[width=7cm]{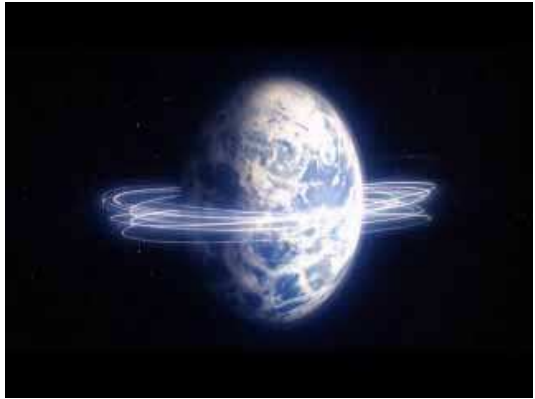}
    \end{center}
    \caption{Superman rotates fast around Earth. Unfortunately the directions of the rotations are
             not seen in this picture. The movie correctly shows Earth rotating from left
             to right as we
             look at this picture, while superman revolves around it from right to left.}
    \label{fig:superman}
    \end{figure}

There are few scenes in all of movies ever produced that  rewrite
so many physics laws as this one does. First of all, why the
director relates the direction of Earth's rotation to the
direction of time is a mystery. Why not the direction of the
Earth's revolution around the Sun? Or maybe, the Moon's revolution
around Earth. Time is a `mystical' quantity in our universe that
is very hard to be explained. Even more, one can make a
distinction about several different kinds of time: fundamental
time, thermodynamic time, etc. Current proposed theories of
unification delve into this topic but the issue is far from being
resolved. However, one thing is very well understood: just
changing of direction of the motion of an object will not do
anything to the flow of time.

Even if the reversal in the direction of Earth's rotation would
reverse the flow of time, why, after  the original direction of
rotation has been restored, the events will not repeat themselves
in the same sequence\footnote{If the events repeat themselves
\emph{identically}, then notice that Superman created a perpetual
loop of events. This loop will be repeated for ever.}? To allow
for a different outcome, we must assume that Superman's action
created a parallel universe that is identical to the universe he
knew up to the point that Lois may or may not be killed. At that
point, Superman's actions send him in one of the universes in
which Lois is not killed.

All the previous discussion is really science fiction. Let's not
pursue it, but instead discuss more down-to-earth flaws of the
scene. As we clearly see in the movie, Superman flies outside
Earth's atmosphere before he starts his revolution. It is easy to
understand how he got there: he got a push from the ground, then
from the air by pushing them in the opposite direction
(action-reaction law). But if he is eventually outside the
atmosphere, how does he propel himself? He cannot get a push from
anything. He could in principle expels mass...which should come
from his own body! Unfortunately, he cannot afford doing
it\footnote{Compute how much of his mass Superman needs to use in
order to match Earth's angular momentum.}. So, despite his good
intentions, what he set out to do cannot  be done, not even by
Superman.

Even so, let's assume that the impossible (propelling himself in
empty space) is possible. The law that the director is using
(actually, is attempting to use) is angular momentum conservation.
Superman and Earth initially have a combined net angular momentum.
If one of them changes its angular momentum, then the other must
change it accordingly such that the sum will remain unchanged.
Superman speeds up, so the Earth must slow down. However, the
director has an incorrect understanding of the law. The movie
shows Superman revolving around Earth in a direction that is
opposite compared to Earth's revolution. This is exactly opposite
scenario of what the director wants: this will increase Earth's
rate of rotation. It is easy to see why. Superman will accelerate
due to a push from Earth. He, of course, applies an opposite push
to Earth. But if he flies in a direction opposite to Earth's
rotation, his push is along Earth's rotation and, therefore, will
speed up Earth's rotation.

Again, let's ignore this `little' detail and assume that Superman
flew in the correct direction. How far away from Earth should he
be? And what should his speed be? Earth is approximately a sphere
of radius $R=6370 km$ and mass $M=6\times10^{24} kg$. The moment
of inertia of a sphere rotating about a diameter equals $I=(2/5) M
R^2$. Therefore its angular momentum is $L_{Earth}=I\omega$ where
$\omega$ is $2\pi$ radians per day or
$\omega=7\times10^{-5}rad/s$. Superman is initially rotating with
Earth's speed:
$$
  L_{superman} \= I_{superman} \omega \= mR^2\omega~.
$$
Since superman's mass (that we assume for simplicity to be about
$100kg$) is negligibly small compared to Earth's mass,
$L_{total}\simeq L_{Earth}$. When Earth comes to rest, only
Superman has angular momentum
$$
  L_{superman}' \= m v d~,
$$
where $v$ is his speed and $d$ his distance from the center of the
Earth. Conservation of angular momentum requires
$$
   m v d \= {2\over 5} M R^2 \omega~,
$$
or
$$
   v d \= 68\times 10^{30}~{m^2\over s}~.
$$
For any distance that is less than $d_0\=2.3\times10^{23}m$, the
required speed is greater than the speed of light $c=300,000km/s$.
Keeping his speed below $c$, implies that he will have to go far
away...further than $d_0$. The universe is about 14 billion years old.
During this time, light has traveled a distance
$1.3\times10^{26}m$. Certainly bigger than what Superman needs.
But does he have time to finish what he started?  He would need to be
moving at nearly the speed of light in a circle whose radius around Earth is
equal to the distance to the edge of the visible universe
when it was 1/1000 of its present size.

\subsection{Impressive Special Effects Imply Impressive Lack of Science Literacy}

In \emph{X-Men: The Last Stand} \cite{X3}, Magneto, the leader of
the brotherhood of X-Men that resists humans performs the
following feat. When his army is ready to attack the island of
Alcatraz where the research institute for the curing of the
X-disease is located, he uses his ability of manipulating magnetic
fields to cut the Golden Gate Bridge loose and relocate it between
San Francisco's port and the island\footnote{Time:
1:13:15--1:16:23.}. The relocation of the bridge gave to the
director an opportunity for great special effects. However, even
with the acceptance of Magneto's special powers, it is an
unrealistic scene given the physical laws in our universe.

    \begin{figure}[h!]
    \begin{center}
    \includegraphics[width=7cm]{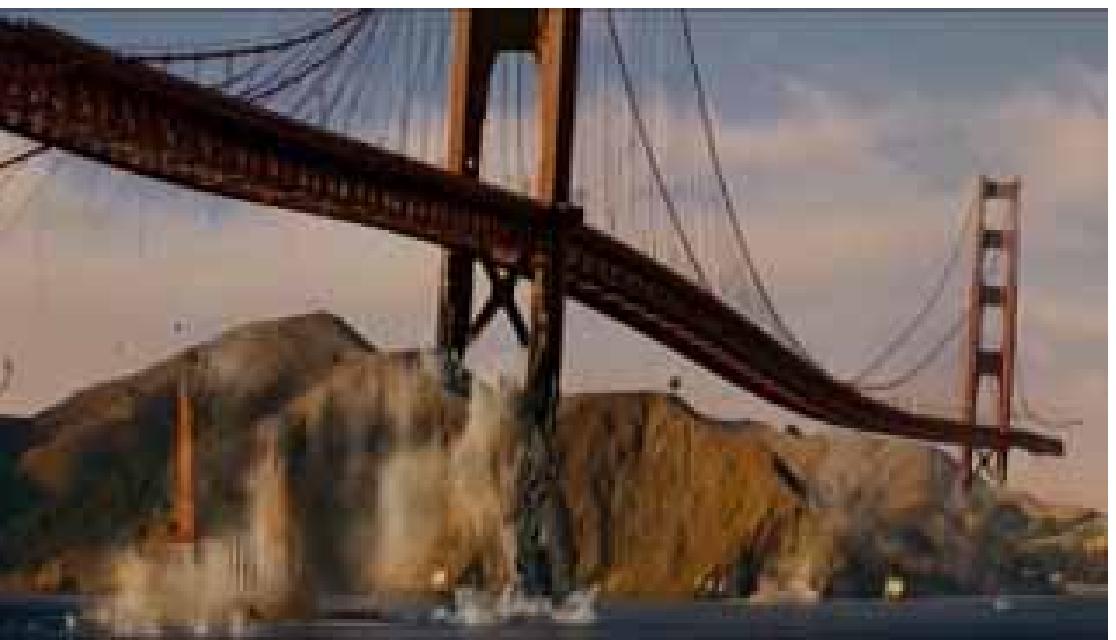}
    \includegraphics[width=7cm]{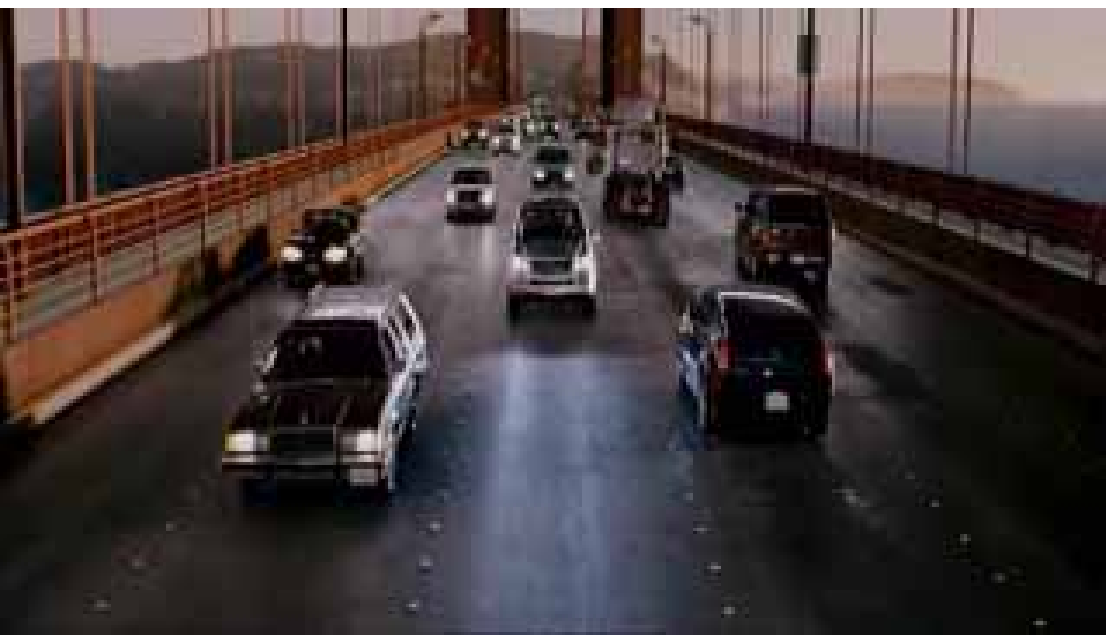}
    \end{center}
    \caption{Left: The section of the Golden Gate Bridge that Magneto transferred.
    Right: The traffic on the bridge before Magneto's attack. }
    \end{figure}

Extensive information about the Golden Gate Bridge can be found at
its website \cite{GoldenGate}. In particular, the mass of Bridge,
not including anchorages and north and south approaches, but
including suspended structure, towers, piers and fenders, bottom
lateral system and orthotropic re-decking is 419,800 short tons or
$380,800,000 kg$. Also, the length of the suspension spans,
including the main span and side spans, is 1,966 meters. Given the
quick shot of traffic on the bridge, we can approximate that when
Magneto cut the bridge, we have about 1 car every 4 meters of the
bridge. This is equal to a total of 393 cars. Since an average car
is about 1,000 kg the total mass of the cars is about 393,000 kg.
We may add the mass of the drivers and passengers in the cars but
we can easily see that the main mass comes from the construction
of the bridge, not the load. So, let's assume that the combined
mass is of the order of 400,000,000 kg.

Using a map we can see that the bridge must be moved about 5400
meters. The movie implies that the whole affair happens within few
minutes. If this means 5 minutes, then the speed should be $18m/s$
or about $65 km/h$. For simplicity, we shall assume a speed of
$10m/s$.

All this would imply kinetic energy of
$$
   K \= {1\over 2} \, m\, v^2 \= 20,000,000,000~J~,
$$
for the transportation of the bridge. Assuming that the bridge is
not lifted  above the ground more than its original clearance,
there is no energy expended for potential energy.
However, in order to break the bridge free work has to be done; we
ignore this since it is harder to compute and adds nothing to the final
conclusion.

The energy for the relocation of the bridge is provided by Magneto
through the magnetic fields he can create. Ultimately the creation
of the magnetic fields originates in the cells of his body that
obtain the energy from the food consumed by Magneto. One Calorie
is 4,200 Joules. Therefore, the energy required for the
transportation of the bridge is equivalent to 4,761,900 Calories.
An average male needs about 2,500 Calories a day just to support
the smooth functioning of his body. Just the task of the
transportation of the bridge requires the prior consumption of
food for 1900 males. Since Magneto has not eaten anything while he
is performing the task, the energy should come at the expense of
his body mass. One pound of fat is about 3,500 Calories. In other
words, Magneto should lose at least 1350 pounds while he
transported the bridge!

Of course, transporting the bridge is by no means Magneto's
only feat. We watch him performing a series of feats,
one after the other. Therefore, the problem is way more serious
than what our calculation shows.

In order to be fair, we must observe that our comments above are
valid only when Magneto's body produces energy through chemical
reactions. If the energy produced is due to nuclear
fusion---exactly the same way energy is produced by the Sun---then
the difficulty we encountered disappears. During nuclear fusion, a
change of mass is observed. The mass that was lost was converted
to energy in accordance to Einstein's celebrated equation:
$$
    E \= \Delta m \, c^2~.
$$
The transportation of the bridge would require  a loss of body
mass of $0.000000225 kg$. This is a very small amount of mass.

In any case, even if Magneto does not have to lose body mass, the
way the situation is presented in the movie is still unrealistic.
The average power of Magneto's body is
$$
   P \= {20,000,000,000~J\over 540~s} \= 37,037,037~Watts~.
$$
In the same way  that the Sun or a light bulb shines when it
produces energy, Magneto should also shine. It is worthwhile to
understand how bright (literally) he should be. The filament of a
standard incandescent $60 W$ light bulb is made of tungsten and it
is about $2 m$ long and about $0.25 mm$. This would imply a
surface area of $0.003m^2$. Then the intensity of the light bulb
is
$$
   I_{bulb} \= {60~W\over 0.003m^2} \= 20,000~{W\over m^2}~.
$$
The surface area of the human body is about $1.8m^2$; let's round
it up to $2m^2$. Then Magnetos's intensity is
$$
  I_{Magneto} \= 18,518,519~{W\over m^2}~.
$$
Magneto's body should be shining about 926 times stronger than a
$60 W$ light bulb!

The scene has additional problems not directly related to Magneto.
A suspension bridge takes its rightful name from the fact that the
load of the bridge is suspended by vertical steel wires hung by
cables\footnote{One can work out easily that, for uniform load,
the shape of these cables must be parabolic. This gives the
suspension bridges their familiar look.} which, in turn, are
secured between two towers. The towers are anchored by additional
cables at the ground. All tension of the bridge and its load
eventually is transferred to the ground through this series of
cables. Magneto, in his attempt to relocate the bridge, cut the
suspension cables of the bridge as the movie so clearly presents.
However, once this is done, there is nothing to support the bridge
with its load and, therefore, it should now collapse\footnote{Of
course, Magneto can choose to create a magnetic field to support
the bridge while it is transported and while in use in its new
location. Of course this must be done in the expense of more
energy from his body. Incidentally, notice that it is not clear
why Magneto chooses to carry his army with the bridge as he
can---more easily---hijack a ship and (magnetically or manually)
drive it to the island.}.

Notice that when Magneto drops the bridge on Alcatraz, he flies
above it\footnote{Magneto's flying actually creates another hole
in the movie's plot. If Magneto has the ability to levitate
humans, why not to just fly his army to the island? An additional
comment may be in order here since the reader may be wondering if
it is possible to levitate humans by the use of magnetic fields.
Actually, it is. All objects, including plastic, wood, and
biological tissue, have  magnetic properties that can be used in
similar ways. However, such objects demonstrate only a very weak
magnetic behavior. Therefore, one needs a large applied field in
order to levitate them. Levitation of small biological objects
(live frog, grasshopper, hazelnut, etc.) was first achieved by a
group of researchers in the Nijmegen High Field Magnet Laboratory
in the Netherlands.}. It is not clear how the audience should
interpret this. Should we assume that he does so to avoid the
impact which is hard enough to create serious damage? Or should we
assume that he does so to show his superiority? Probably, the
latter since the director seems not to understand what the effect
from the drop may be on the objects siting on the bridge. None of
the cars on the bridge, nor anyone of Magneto's army seem to have
been affected the slightest by the fall. An interesting problem
for the reader might be the estimation of the force acting on the
objects as a result of the fall of the bridge.

\subsection{Artistic Exaggerations that Lead to the Ridiculous}

In \emph{The Chronicles of Riddick} \cite{Riddick} there is a
planet called Crematoria. It takes its name from the harsh
environment it offers. As a spaceship approaches Crematoria, a
quick close-up of the `Course Plotted' panel is shown. There we
read the temperature differential\footnote{From the close-up we do
not know if the temperature is measured in degrees Celsius or
Fahrenheit.} on the two sides of the planet, $-295^\circ$F and
$+702^\circ$F, while  a crew member states loud ``700 degrees on
the day side; 300 below on the night side"\footnote{Time:
0:53:39--0:53:56.}.

Riddick, after being imprisoned in an underground prison in
Crematoria, succeeds in escaping with a group of other prisoners
just before sunrise. The group then starts to run towards a
spaceship (which fortunately happens to be in the right direction)
away from the coming sunshine that brings the devastating daylight
temperatures. The race is hard. As they climb a cliff, Riddick's
old friend Kyra falls behind and the morning heat traps her behind
a rock. Unable to leave Kyra face her fate, Riddick tries his
ultimate trick: he drops the water from a flask on himself,  ties
a rope on the top of the cliff,  swings and...saves Kyra. After
the rescue, we still see the vapors from the water (and his sweat
supposedly)\footnote{Time: 1:29:47--1:32:27.}.

    \begin{figure}[h!]
    \begin{center}
    \includegraphics[width=7cm]{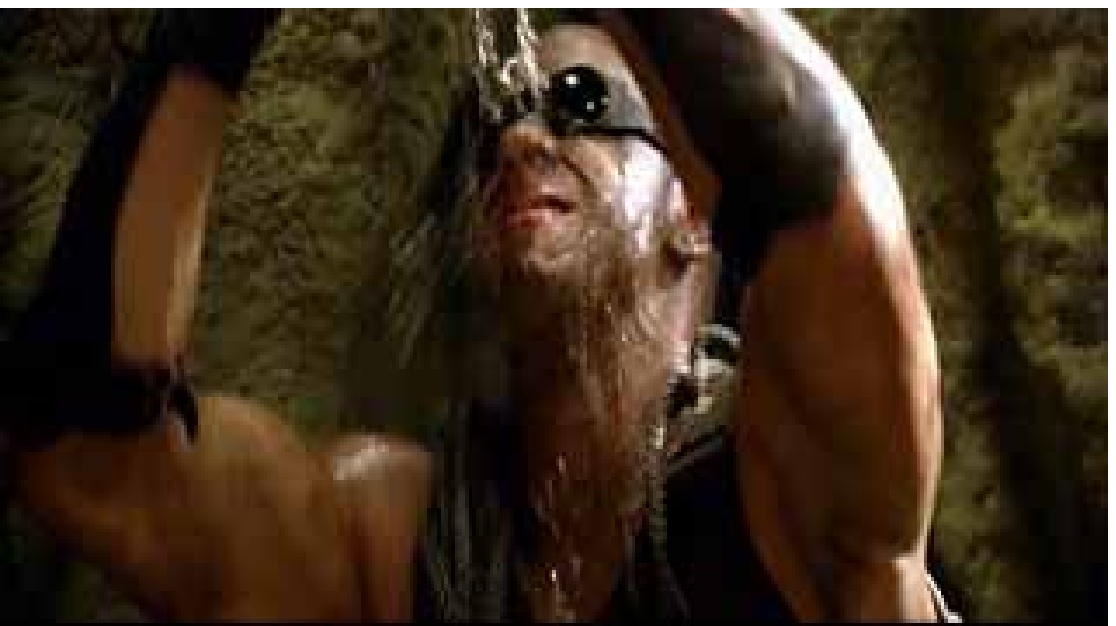}
    \includegraphics[width=7cm]{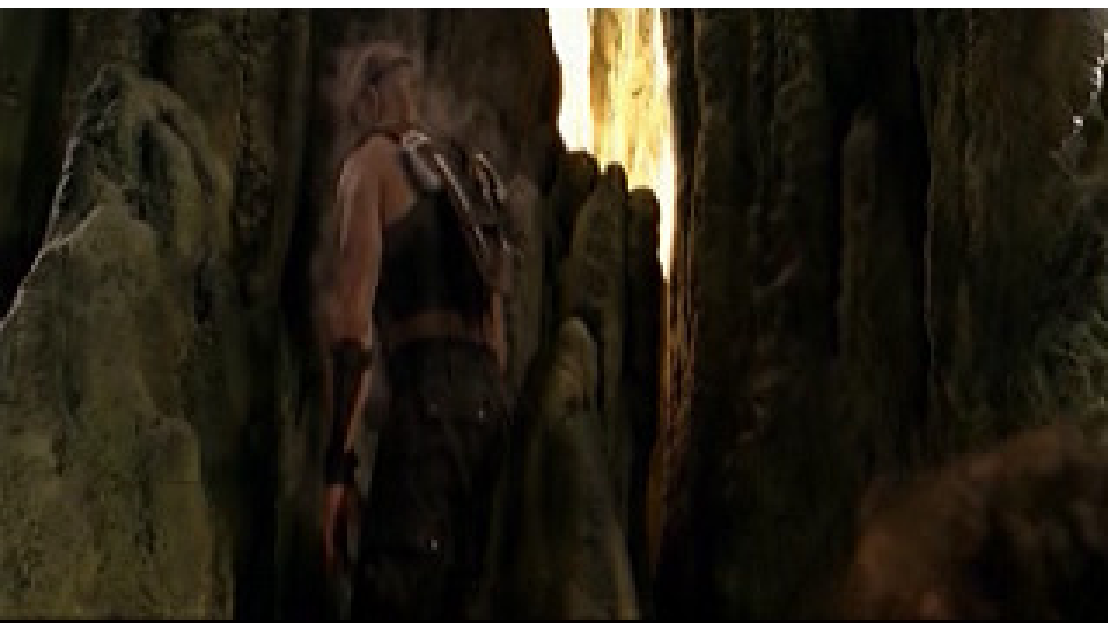}
    \end{center}
    \caption{Left: Riddick drops a flask of water on himself
             before attempting the rescue of Kyra.
             Right: Vapors of water are seen around the body of Riddick after the rescue of
              Kyra.}
    \end{figure}

The reader after watching the scene will certainly have realized
that, scientifically, it is rubbish. First of all, why are the
escaped prisoners only afraid of the incoming sunshine?  Why
aren't they also afraid of the dark side? Isn't the temperature in
the dark side $-295^\circ$F?  Well, as always, let's give the
director the benefit of the doubt. Let's pretend that due to the
incoming heat and the fact that the heroes are caught exactly in
the middle of dark-light, the temperature they experience is
something that they can tolerate.

At the heart of the scene is the harshness of the environment on
Crematoria. The director has tried hard to make the audience
appreciate how harsh the conditions are. He shows that, as the
daylight comes, the `temperature differential' creates a heat wave
that incinerates anything it finds in front of it. One of the
escaped prisoners attempted to look at Kyra's rescue and was
incinerated by the wave in seconds. After all this effort to
present an impossible situation, the director then shows to the
audience that a flask of water is the solution... This is worth of
some discussion.

    \begin{figure}[h!]
    \begin{center}
    \includegraphics[width=4cm]{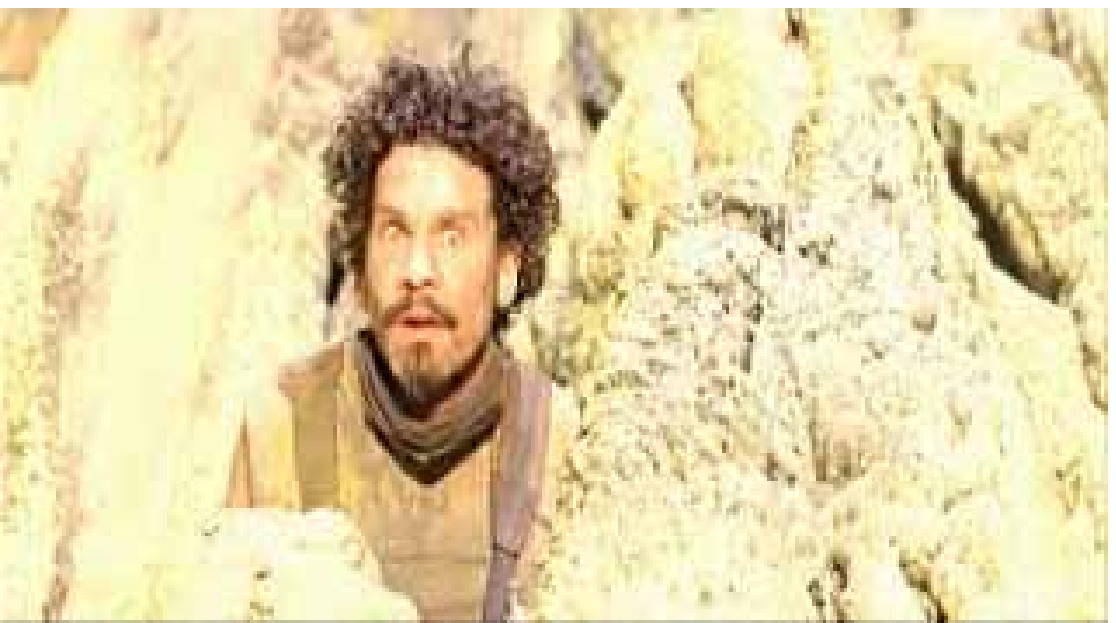}
    \includegraphics[width=4cm]{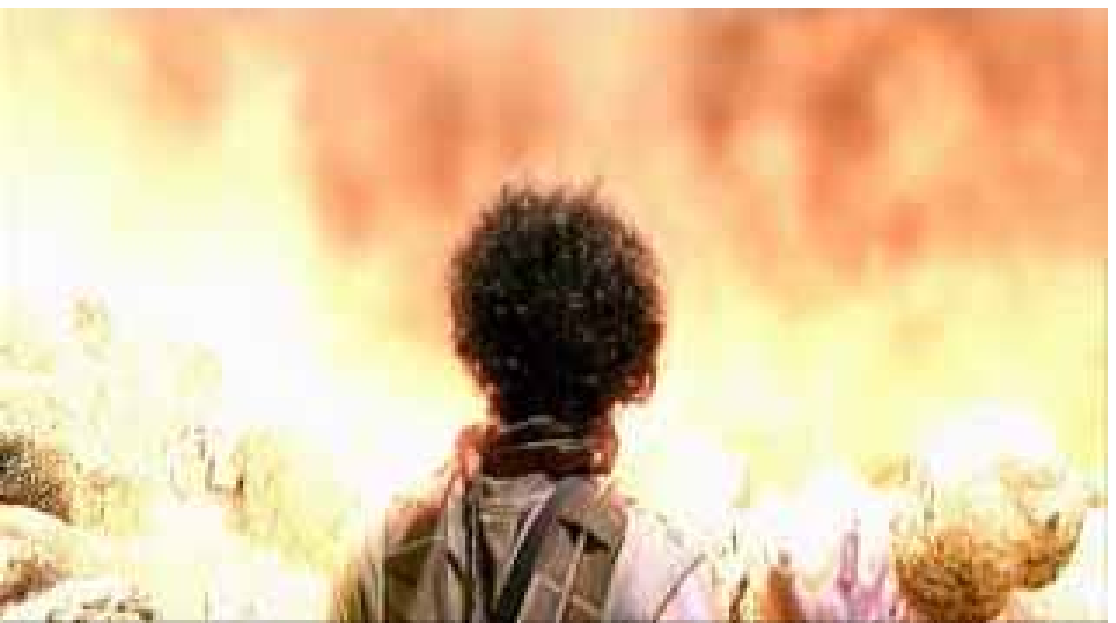}
    \includegraphics[width=4cm]{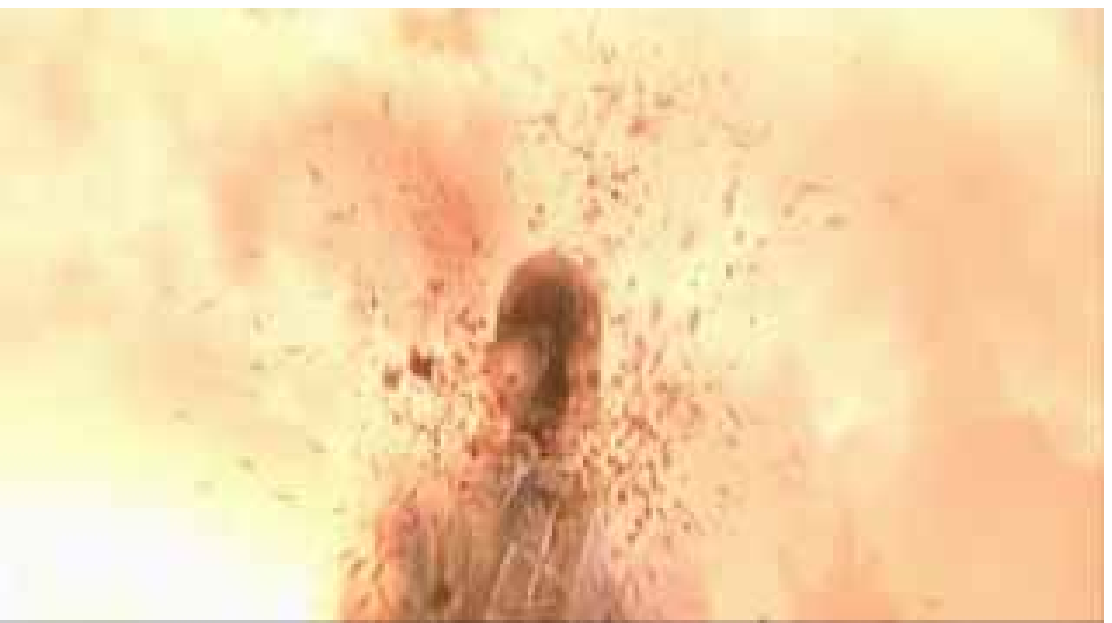}
    \end{center}
    \caption{A sequence of stills from the incineration of one the escapees.
             The incineration happens in seconds.}
    \end{figure}

Actual cremation takes place at an initial temperature of
$700^\circ$C ($1290^\circ$F) which increases to
$900^\circ-1100^\circ$C ($1650^\circ-2010^\circ$F) during
combustion. The cremation lasts 60-90 minutes for obese people and
up to 120 minutes for thin and underweight people. The temperature
on Crematoria is way below the required temperature for
incineration. It simply cannot happen and if it could, the time
shown for it is nonsense. In the piece \emph{Special Effects
Revealed} of the DVD bonus features, Peter Chiang, Visual Effects
Supervisor, explains that the original script, written by David
Twohy, was using degrees Celsius. It would be nice if a simple
error would be a solution to this problem. With this point of
view, the temperature is  enough to start the cremation but still
below that required at the later stage. It would require,
 at least, the full time of cremation to get the ashes of the incinerated
 people. However, there is an even worse problem in this interpretation.
The temperature $-300^\circ$C does not exist! The lowest
temperature in our universe is that of the absolute zero and this
is $-273^\circ$C.

Although the temperature is not high enough to get spontaneous
incinerations, it is too high for survival. Our heroes should be
fried. At such intense temperatures, the moisture of the skin will
evaporate quickly and the dry skin will be severely damaged. Let's
look at the water Riddick dropped on himself. From the size of the
flask, we must assume that it was not more than 1 liter
(equivalently 1 kilogram in mass)---in fact this amount is
probably an exaggeration. Most of the water is lost; very little
will adhere to his skin. We will use his head as example for the
calculation. Probably no more than $10g$ of water will remain on
his head. Say that the water was absorbed uniformly by the head.
If we approximate the surface area of the head to be $0.2m^2$ then
the water density\footnote{It is actually an advantage that
Riddick has shaved his head. Hair might retain a little more water
but it will have way higher surface area leading to faster
evaporation.} covering Riddick's head is $\sigma=0.05kg/m^2$. We
further assume that the water was at room temperature---say about
$25^\circ$C. If $L=2,257 kJ/kg ~^\circ$C is the latent heat of
vaporization of water and $c=4.2 kJ/kg~^\circ$C is the specific
heat of water, then the total energy per unit area $\epsilon$
required to evaporate the water is:
$$
  \epsilon \= \sigma (c \Delta T +  L) ~,
$$
 or  about $129 kJ/m^2$.
To find out how long it takes for this water to evaporate, we must
approximate the
 intensity of the Sun's energy output on Crematoria.
 We can get a rough estimation about this energy by looking at
 Mercury in our solar system. Mercury has a temperature
 differential similar to Crematoria, $-280^\circ$F to
 $800^\circ$F. There are several notable differences however.
 Crematoria seems to have a rotation, a gravitational field, and
 an atmospheric pressure and content equal (or at least close) to
 those of Earth. This is not the case for Mercury. In principle,
 it is hard to understand how Crematoria could have maintained an atmosphere
 due to its proximity to the Sun. The solar wind should have
 washed the atmosphere out of the planet as it has been done on
 Mercury. However, the movie shows no worries about this, so we won't worry about it either.
 The solar wind would have made a better special effect compared
 to the temperature differential used. In any case, we can
 hypothesize that, for some unknown reason, Crematoria has been able to maintain
 its atmosphere. Of course, having an atmosphere will create
 meteorological phenomena that will affect the temperature on the planet. All
 this indicates that the assumption that we can, as a rough estimate, equate the energy
that reaches Crematoria to the energy that reaches Mercury. In any case, in astronomical units, Earth's
mean distance from the Sun is $d_{Earth}=$1AU and Mercury's mean
distance is $d_{Mercury}=0.306$AU. Also, the intensity of sunlight
on Earth, just outside the atmosphere, is $I_{Earth}=1.4kW/m^2$.
All this would imply a sunlight intensity at the surface of
Mercury equal to
$$
   I_{Mercury}\= I_{Earth}\,\left({d_{Earth}\over d_{Mercury}}\right)^2
   \= 14.95{kW\over m^2}~.
$$
Therefore the sun delivers about $15 kJ/m^2$ per second and
implies that the protective layer of water spread by Riddick would
last less than 9 seconds under direct exposure to the sunlight. Of
course, because the rising heat in the air, as soon as Riddick
dropped the water on him, the water started to absorb heat. This
reduces the time this protective layer needs for evaporation.
Hopefully, he gains enough time from his sweat to finish his task
otherwise his skin will get cooked...

\section{Conclusions}

Hollywood directors and special effects creators work hard to
create impressive scenes in movies to excite the audience.
However, many scenes are created with absolute disregard of the
physical laws in our universe. Sometimes the scene is so
profoundly wrong that it is hard to be missed. However, often the
absurdity is hard to detect by people not very fluent in science
literacy and  untrained in critical thinking. In this way,
Hollywood is reinforcing (or even creating) incorrect scientific
attitudes that can have negative results for the society. This is
a good reason to recommend that all citizens be taught critical
thinking and be required to develop basic science and quantitative
literacy.

\section*{Acknowledgements}

We thank Prof. Dr. Rainer M\"uller for inviting us to write the
present article for Praxis der Naturwissenschaften Physik. We also
thank Jared Gottesman for proofreading the article and providing
us feedback.


\end{document}